\documentclass[conference,letterpaper]{ieeeconf}
\IEEEoverridecommandlockouts
\usepackage{textcomp}
\usepackage{microtype}
\usepackage{graphicx}
\graphicspath{{figures/}}
\usepackage[font=footnotesize,format=hang,labelfont=bf]{caption}
\usepackage{subcaption}
\input{styleguides/tikzsetup.sty}
\usepackage[english]{babel}
\usepackage{amsmath,amssymb,amsfonts,mathtools}
\usepackage{array}
\usepackage{xspace}

\usepackage{styleguides/CommonMacros}

\newcommand{\ts}{t_{\mathrm{s}}}
\newcommand{\epsk}{\varepsilon_k}
\newcommand{\epsbar}{\bar{\varepsilon}}
\newcommand{\epsdes}{\varepsilon_{\mathrm{des}}}
\newcommand{\Adnom}{A_{\mathrm{d},0}}
\newcommand{\Adk}{A_{\mathrm{d},k}}
\newcommand{\Bdnom}{B_{\mathrm{d},0}}
\newcommand{\Bdk}{B_{\mathrm{d},k}}
\newcommand{\DeltaA}[1]{\Delta A(#1)}
\newcommand{\DeltaB}[1]{\Delta B(#1)}
\newcommand{\Dk}{\Delta_k}

\newenvironment{ssm}[1]
{
  #1: \; \left\{
  \begin{array}{rcl}
}
{
  \end{array} \right.
}

\begin{document}

\title{Robust Control Design\\for LTI Systems with Sampling-Time Jitter%
\thanks{This work has been submitted to the IEEE for possible publication. Copyright may be transferred without notice, after which this version may no longer be accessible.}}

\author{\authorblockN{Dieter Schwarzmann}\authorblockA{Institute for Systems Theory and Automatic Control,\\
University of Stuttgart, Stuttgart, Germany\\
dieter.schwarzmann@ist.uni-stuttgart.de}
\and
\authorblockN{Jan Lunze}\authorblockA{Faculty of Electrical Engineering and Information Technology,\\
Ruhr-University Bochum, Bochum, Germany\\
Lunze@atp.rub.de}}

\maketitle

\begin{abstract}
Sampling-time jitter makes a sampled-data control loop time-varying, so standard LTI robust-control synthesis does not apply directly. This paper reformulates dynamic output-feedback design under bounded, unmeasured sample-to-sample jitter as a robust-control problem. The exact zero-order-hold model is decomposed into a nominal time-invariant plant and a memoryless, time-varying uncertainty block with a norm bound representing the sampling-time jitter. As the block is memoryless, the operator small-gain condition for the time-varying loop reduces to a standard robustness test with a matrix-norm bound. It is shown that the same test also guarantees uniform exponential stability for every admissible jitter sequence. Embedding the decomposition into a generalized plant with performance channels serves two synthesis objectives with standard robust-control tools: Robust stability with optimized nominal performance through structured $\Hinf$ synthesis, and robust performance through $\mu$ synthesis with constant scalings. A numerical example shows that jitter can destabilize a control loop with a nominal $\Hinf$ controller, whereas both robust designs remain stable and the robust-performance design is less sensitive to jitter at the cost of nominal performance.
\end{abstract}

\section{Introduction}

\begin{figure}[htbp]
  \centering
  \includegraphics[width=4.8cm]{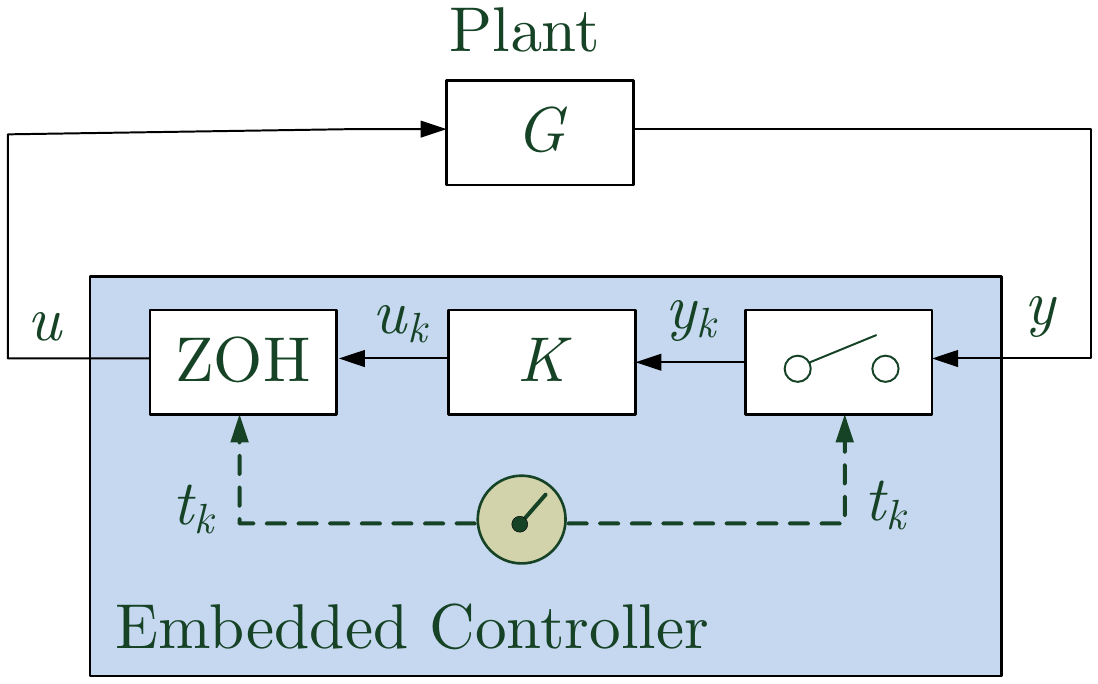}
  \caption{Embedded control loop with controller $K$ (reference and error summation omitted). Sampling and ZOH share the instants $t_k$.}
  \label{fig:embedded-loop}
\end{figure}

Embedded controllers that share resources with other tasks get preempted and can miss their nominal sampling period~$\ts$ (\figref{fig:embedded-loop}). Sampler, controller update, and zero-order hold (ZOH) act at the same instants $t_k$, with varying interval
\[
  \Delta t_k=t_{k+1}-t_k=\ts(1+\epsk),\qquad|\epsk|\leq\epsbar<1,
\]
where $\epsk$ is the relative jitter and $\epsbar$ the \emph{jitter bound}. This timing situation belongs to the classification in our earlier paper \cite{IFAC2026}. The controller cannot measure\footnote{The length of the current interval is known only once it is over.} $\epsk$, and we make no stochastic assumption about its sequence.

Jitter makes the sampled plant time-varying, so nominal LTI design gives no robustness guarantee. We seek an LTI discrete-time output-feedback controller that stabilizes the loop for every sequence in the \emph{jitter envelope}
\begin{equation}\label{eq:envelope}
  \mathcal{E}(\epsbar)=\bigl\{\{\varepsilon_k\}_{k\geq0}\;\big|\;|\epsk|\leq\epsbar\ \text{for all}\ k\bigr\}
\end{equation}
with optimized nominal performance or with a guaranteed performance bound. The jitter bound limits each interval, while the envelope $\mathcal{E}(\epsbar)$ collects the \emph{admissible} sequences.

We reformulate the time-varying design problem as a norm-bounded robust-control problem. The exact ZOH decomposition into nominal LTI dynamics and a memoryless jitter block is known \cite{Fujioka2009TAC,Fujioka2008,FujiokaNakai2010,KaoFujioka2013}. We apply the operator small-gain theorem as a robustness test for this feedback loop and  prove that this test reduces to a matrix-norm condition on the memoryless jitter block. This condition ensures stability for every jitter sequence in~$\mathcal{E}(\epsbar)$. Combining this certificate with performance channels in a generalized plant leads to methods for fixed dynamic output-feedback synthesis that use standard robust-control tools\footnote{By standard robust-control tools we mean the off-the-shelf commands of MATLAB's Robust Control Toolbox: \texttt{hinfsyn} for nominal $\Hinf$ synthesis, \texttt{systune} for constrained structured $\Hinf$ synthesis, and \texttt{musyn} for $\mu$ synthesis, or the equivalent commands of other toolboxes.}.

The paper presents four main results:
\begin{itemize}
  \item a generalized plant that combines the exact jitter decomposition with performance channels and dynamic output feedback in the P--K--$\Delta$ form used by standard robust-control tools (Section~\ref{sec:stability-guarantee});
  \item an LTI-computable small-gain test that certifies finite-gain stability and uniform exponential stability for every admissible jitter sequence, including the inter-sample plant state (Section~\ref{sec:stability-guarantee});
  \item a synthesis workflow for robust stability with optimized nominal performance and for robust performance with the constant-scaling settings needed to retain certificates for sample-varying jitter (Section~\ref{sec:tooling});
  \item the jitter bound $\epsbar$ as an explicit design parameter, which lets the engineer trade conservatism for nominal performance, and a numerical comparison showing how nominal, robust-stability, and robust-performance designs respond to jitter (Section~\ref{sec:example}).
\end{itemize}

From a practical viewpoint, the main result is the design workflow of Section~\ref{sec:tooling} in which off-the-shelf robust-control tools return provably correct controllers for all jitter sequences of \eqref{eq:envelope}.

\section{Related Work}

The closest precedent of this paper is Fujioka's exact discrete-time embedding of aperiodic sampling: Stability analysis for a given state-feedback gain \cite{Fujioka2009TAC} or dynamic output-feedback controller \cite{Fujioka2008}, followed by state-feedback synthesis with dedicated LMI algorithms \cite{FujiokaNakai2010}. Fujioka and Nakai~\cite{FujiokaNakai2010} identify output-feedback synthesis and induced-norm performance, including inter-sample behavior, as future directions. This paper extends the embedding to fixed discrete-time dynamic output-feedback synthesis with sampled-signal performance specifications, using standard robust-control tools with certificates valid for unmeasured, sample-varying jitter.

Hetel et al.~\cite{Hetel2017} survey other approaches. Hybrid-system and input-delay methods give dedicated stability and synthesis conditions \cite{Naghshtabrizi2006,Fridman2004}. Suplin et al.~\cite{Suplin2007} synthesize sampled-data $\Hinf$ output-feedback controllers for unknown bounded intervals, but their controller states evolve continuously between samples, whereas ours follow a fixed discrete-time recursion. Clock-dependent designs in \cite{Briat2013} use elapsed sampling time, whereas Markov jump models in \cite{Mustafa2012} require a probability law. Our approach requires neither measured jitter nor a probabilistic model.

\section{Discrete-Time Control Loop with Jitter}

\subsection{The Continuous-Time Plant and Its Nominal Discretization}

The continuous-time LTI plant $G$ is:
\begin{equation}\label{eq:G_definition}
\begin{ssm}{G}
    \dot{x}(t) &=& A x(t) + B u(t),
     \quad x(0)=x_0, \\
    y(t) &=& C x(t),
\end{ssm}
\end{equation}
where $A \in \mathbb{R}^{n \times n}$, $B \in \mathbb{R}^{n \times m}$, $C \in \mathbb{R}^{p \times n}$. We omit a direct feedthrough $y=Cx+Du$ to simplify notation, but all results also hold for $D\neq 0$. At the reference interval $\ts>0$, the exact ZOH equivalent of $G$ is
\begin{equation}\label{eq:nominal-discretization}
  \begin{ssm}{G_{\mathrm{d},0}}
    x_{k+1} &=& \Adnom x_k + \Bdnom u_k,\\
    y_k &=& C x_k,
  \end{ssm}
\end{equation}
with $x_{k=0}=x_0$ and matrices
\begin{equation}
\Adnom = e^{A \ts}, \qquad \Bdnom = \int_0^{\ts} e^{A\tau} B \, d\tau.
\end{equation}

\subsection{The Jittered Model}

Let $t_0=0$ and define $t_{k+1}=t_k+(1+\epsk)\ts$. The dimensionless relative jitter satisfies $|\epsk|\leq\epsbar<1$, so positive values lengthen the interval and negative values shorten it. To make the bound symmetric merely simplifies notation. All results hold for an asymmetric range $-1<\varepsilon_{\min}\le\epsk\le\varepsilon_{\max}$ if the maximum in \eqref{eq:uncertainty-bound} and the envelope are taken over this range and $\epsbar:=\max(-\varepsilon_{\min},\varepsilon_{\max})$ elsewhere. Because $\epsk$ is unmeasured and unmodeled, certificates must hold for every sequence in the envelope $\mathcal{E}(\epsbar)$ defined in \eqref{eq:envelope}. The exact ZOH model is
\begin{equation}\label{eq:jittered-discretization}
  \begin{ssm}{G_{\mathrm{d},k}}
    x_{k+1} &=& \Adk\, x_k + \Bdk\, u_k,\\
    y_k &=& C x_k
  \end{ssm}
\end{equation}
with $x_{k=0}=x_0$ and matrices
\begin{equation}
\Adk = e^{(1+\epsk) A \ts}, \qquad \Bdk = \int_0^{(1+\epsk) \ts} e^{A\tau} B \, d\tau.
\end{equation}
In \eqref{eq:jittered-discretization}, $x_k=x(t_k)$ and $y_k=y(t_k)$ are the exact state and output at the nonuniform physical instants. The index $k$ counts samples. The dependence on $\epsk$ makes the recursion~\eqref{eq:jittered-discretization} time-varying.

\subsection{Decomposition of the Discrete-Time Model}
\label{Sec:Decomp}
Subtracting the nominal matrices $\Adnom,\Bdnom$ from the jittered ones gives 
\[
   \Adk=\Adnom+\DeltaA{\epsk}, \;\; 
   \Bdk=\Bdnom+\DeltaB{\epsk},
\]
where
\begin{equation}
\begin{aligned}
\DeltaA{\epsk} &:= e^{(1+\epsk) A \ts} - e^{A \ts}\\
  &= \bigl(e^{\epsk A \ts} - I\bigr) \Adnom,\\
\DeltaB{\epsk} &:= \Bdk-\Bdnom
 = \int_{\ts}^{(1+\epsk) \ts} e^{A\tau} B \, d\tau.
\end{aligned}
\end{equation}
The state equation becomes
\begin{equation}
x_{k+1} = \Adnom\, x_k + \Bdnom\, u_k + \DeltaA{\epsk}\, x_k + \DeltaB{\epsk}\, u_k.
\end{equation}
The jitter-driven perturbation is $\DeltaA{\epsk}x_k+\DeltaB{\epsk}u_k$. Both matrices depend only on the current $\epsk$ and carry no internal dynamics, so the perturbation acts as a static gain between the stacked signals $x_k$ and $u_k$. This \emph{static} property lets us bound the perturbation gain offline and certify the loop with a small-gain theorem (Section~\ref{sec:stability-guarantee}) without tracking how jitter evolves in time.
\begin{figure}[htbp]
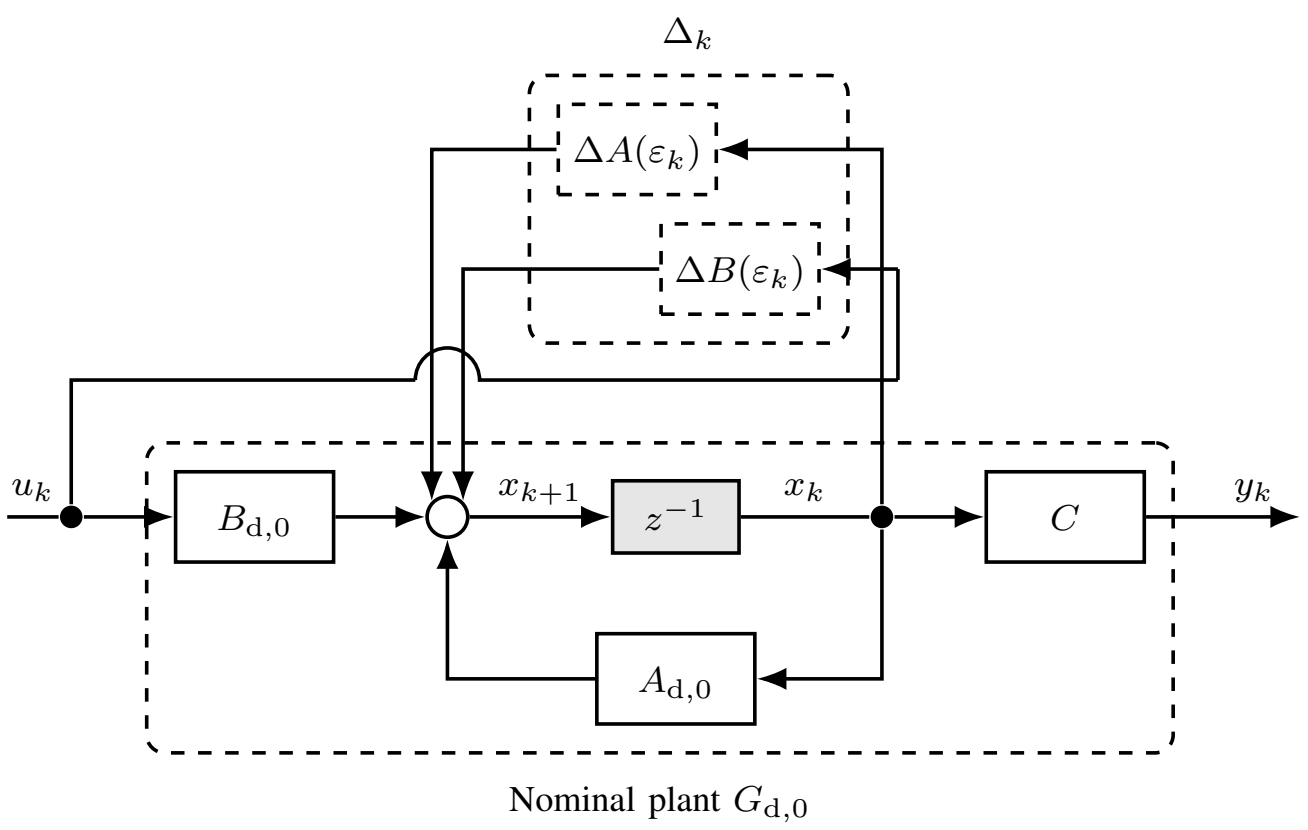

  \centering
  \resizebox{\columnwidth}{!}{%
  \begin{ctrldiagram}[node distance=1.0cm and 1.2cm]
    \node[input]                          (in)    {};
    \node[branch,  right=0.4cm of in]     (bpu)   {};
    \node[block,   right=0.7cm of bpu]    (Bd)    {$\Bdnom$};
    \node[sum,     right=0.7cm of Bd]     (sumx)  {};
    \node[block-delay, right=1.1cm of sumx]   (delay) {$z^{-1}$};
    \node[branch,  right=1.0cm of delay]  (bpx)   {};
    \node[block,   right=0.7cm of bpx]    (Cout)  {$C$};
    \node[output,  right=1.2cm of Cout]   (out)   {};

    \draw[connector] (in)    -- node[signal-h] {$u_k$}      (bpu);
    \draw[signal]    (bpu)   --                              (Bd);
    \draw[signal]    (Bd)    --                              (sumx);
    \draw[signal]    (sumx)  -- node[signal-h] {$x_{k+1}$}  (delay);
    \draw[connector] (delay) -- node[signal-h] {$x_k$}      (bpx);
    \draw[signal]    (bpx)   --                              (Cout);
    \draw[signal]    (Cout)  -- node[signal-h, pos=0.7] {$y_k$}      (out);

    \node[block, below=0.6cm of delay]   (Ad)    {$\Adnom$};

    \draw[signal, vh] (bpx)     to                          (Ad.east);
    \draw[signal, hv] (Ad.west) to                          (sumx.south);

    \node[block-dashed, above=2.2cm of delay, xshift=-0.3cm]  (DeltaA) {$\DeltaA{\epsk}$};

    \draw[signal, vh] (bpx)         to  (DeltaA.east);
    \draw[signal, hv] (DeltaA.west) to  (sumx.north west);

    \node[block-dashed, below=0.2cm of DeltaA, xshift=0.8cm]  (DeltaB) {$\DeltaB{\epsk}$};

    \coordinate (bpuup)         at ($(bpu |- Bd.north)+(0,0.7cm)$);
    \coordinate (DeltaBright)   at ($(DeltaB.east)+(0.6cm,0)$);
    \coordinate (DeltaBrightup) at ($(DeltaBright |- bpuup)$);
    \coordinate (DeltaBmid)     at ($(DeltaBright |- DeltaB.center)$);

    \coordinate (cross_left)    at ($(sumx |- bpuup) + (-0.25cm, 0)$);

    \draw[connector] (bpu) -- (bpuup) -- (cross_left);
    \draw[connector] (cross_left) arc (180:0:0.25cm) -- (DeltaBrightup);

    \draw[connector] (DeltaBrightup) -- (DeltaBmid);
    \draw[signal]    (DeltaBmid) -- (DeltaB.east);

    \draw[signal, hv] (DeltaB.west) to (sumx.north east);

    \node[group-box, fit=(Bd) (delay) (Cout) (Ad)] (Nominalgroup) {};
    \node[below=2pt of Nominalgroup] {Nominal plant $G_{\mathrm{d},0}$};

    \node[group-box, fit=(DeltaA) (DeltaB)] (Deltagroup) {};
    \node[above=2pt of Deltagroup] {$\Dk$};

  \end{ctrldiagram}}
  \caption{Jittered plant $G_{\mathrm{d},k}$: Nominal part (solid) and jitter-dependent perturbations (dashed), grouped into the uncertainty block $\Dk$.}
  \label{fig:Gd-jitter}
\end{figure}

\figref{fig:Gd-jitter} separates the nominal LTI dynamics from the jitter-dependent perturbation channels. Performance channels are added in the generalized plant of \figref{fig:struct1-ports}.

\subsection{Jitter Uncertainty Block}
\label{sec:uncertainty-structures}
The P--K--$\Delta$ form requires the uncertainty $\Delta$ as a block with defined input and output ports. We choose a static, non-square full block acting on the stacked port $[x_k\Tr,u_k\Tr]\Tr$:
\begin{equation}\label{eq:jitter-block}
\Dk: \;\; \Delta(\epsk) := \bigl[\DeltaA{\epsk} \;\big|\; \DeltaB{\epsk}\bigr] \in \mathbb{R}^{n \times (n+m)}
\end{equation}
Equation~\eqref{eq:jitter-block} names the block $\Dk$ and the matrix $\Delta(\epsk)$ that describes it. At sample $k$ the block acts by multiplication with $\Delta(\epsk)$. The ports are
\begin{equation}
\begin{aligned}
u_{\Delta,k} &= \begin{pmatrix} x_k \\ u_k \end{pmatrix},\\
y_{\Delta,k} &= \Delta(\epsk)\,u_{\Delta,k} = \DeltaA{\epsk}x_k + \DeltaB{\epsk}u_k.
\end{aligned}
\end{equation}
This representation needs no factorization of $\DeltaA{\epsk}$ or $\DeltaB{\epsk}$, in particular no inverse of $A$. The packaging is not unique. Other factorizations and structured embeddings exist beyond this scope.\\[1ex]

\noindent
\textit{Offline norm bound.} For each admissible jitter sequence, define the memoryless \emph{jitter operator} $\mathcal D$ on $\ell_2$ by
\begin{equation}
\label{eq:jitter-operator}
   (\mathcal D u_\Delta)_k=\Delta(\epsk)\,u_{\Delta,k}.
\end{equation}
Its induced gain is bounded by~$\bar\delta$:
\begin{equation}\label{eq:uncertainty-bound}
  \begin{aligned}
    \|\mathcal D\|_{\ell_2\to\ell_2}
      &= \sup_{k\geq0}\|\Delta(\epsk)\|_2
      \leq \max_{|\varepsilon|\leq\epsbar}\bar\sigma\bigl(\Delta(\varepsilon)\bigr)
      =: \bar\delta .
  \end{aligned}
\end{equation}
In \eqref{eq:uncertainty-bound}, $\bar\sigma$ is the maximum singular value and the scalar $\bar\delta$ the \emph{uncertainty bound}. The bound depends only on $A$, $B$, $\ts$ and the jitter bound $\epsbar$, not on the realized jitter sequence. One maximization over a grid of jitter values $|\varepsilon|\leq\epsbar$ therefore serves every admissible sequence and every controller designed afterward. Appendix~\ref{sec:analytical-bounds} gives a closed-form bound that needs no grid.

The full block \eqref{eq:jitter-block} discards the one-parameter dependence on $\varepsilon$ and is therefore conservative. Because $u_\Delta$ stacks state and input, its normalization also matters: Port rescaling changes both $\bar\delta$ and the nominal return map. The engineer therefore compares scalings by the resulting small-gain product, not by $\bar\delta$ alone.

\section{Stability Guarantee and Its Interpretation}
\label{sec:stability-guarantee}

\subsection{The Feedback Loop under Consideration}

\begin{figure}[htbp]
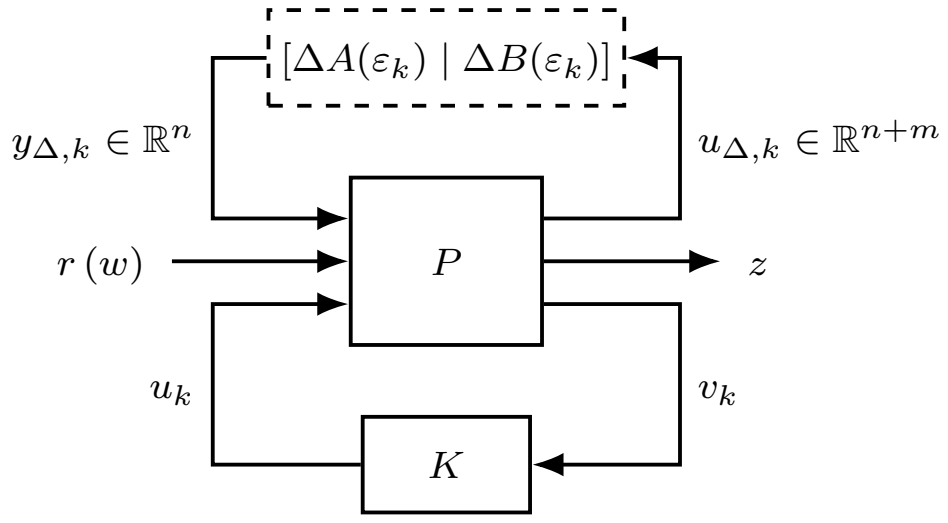

  \centering
  \scalebox{0.9}{%
  \begin{ctrldiagram}[node distance=1.2cm and 2.5cm]
    \node[block, minimum height=3.5em, minimum width=4em]       (P)     {$P$};
    \node[block-dashed, above=0.5cm of P, minimum width=7em]  (Delta) {$[\DeltaA{\epsk} \mid \DeltaB{\epsk}]$};
    \node[block, below=0.5cm of P] (K) {$K$};
    \blockports{P}{3}{3}
    \blockhighways{P}
    \draw[signal, wrap-right=P] (P out 1)    to node {$u_{\Delta,k} \in \mathbb{R}^{n+m}$} (Delta.east);
    \draw[signal, wrap-left=P]  (Delta.west) to node {$y_{\Delta,k} \in \mathbb{R}^{n}$}   (P in 1);
    \extin [-1.3,0]{w}{P in  2}{}
    \extout[1.3,0]{z}{P out 2}{}
    \node[left=2pt of w]  {$r\,(w)$};
    \node[right=2pt of z] {$z$};
    \draw[signal, wrap-right=P] (P out 3) to node {$v_k$} (K.east);
    \draw[signal, wrap-left=P] (K.west) to node {$u_k$} (P in 3);
  \end{ctrldiagram}}
  \caption{Closed loop with the jitter uncertainty block~$\Dk$ and dynamic output-feedback controller~$K$. For the tracking problem the exogenous port carries the reference, $w=r$, the controller input is $v=r-y$, and $z=W_zv$.}
  \label{fig:struct1-ports}
\end{figure}

We synthesize a fixed discrete-time dynamic output-feedback controller
\begin{equation}\label{eq:controller}
\begin{ssm}{K}
    x_{K,k+1} &=& A_Kx_{K,k}+B_Kv_k, \\
    u_k &=& C_Kx_{K,k}+D_Kv_k.
\end{ssm}
\end{equation}
Its matrices are constant in $k$ and independent of the actual sampling interval. For tracking with exogenous input $w=r$, the nominal generalized plant is
\begin{equation}\label{eq:tracking-generalized-plant}
\begin{ssm}{P}
    x_{k+1} &=& \Adnom x_k+\Bdnom u_k+y_{\Delta,k},\\
    u_{\Delta,k} &=& \begin{pmatrix}x_k\\u_k\end{pmatrix},\\
    v_k &=& r_k-y_k,\quad y_k=Cx_k,\\
    z &=& W_zv,
\end{ssm}
\end{equation}
where $W_z$ is a fixed discrete-time LTI weighting filter and $v_k$ is the tracking error fed to $K$. The inputs of $P$ are $[y_\Delta,r,u]$ and its outputs are $[u_\Delta,z,v]$. The exogenous input of $P$ is the reference, denoted $w$ in the closed-loop description below, so that $w_k=r_k$ holds throughout. The full state $x_k$ appears only at the uncertainty port $u_{\Delta,k}$, which models how jitter acts on the plant. The controller $K$ receives only the measured output $y_k$ through $v_k$, so it needs neither the state nor a jitter measurement. Connecting $\Delta_k$ recovers the sampled plant dynamics. Other disturbance and performance channels can be added similarly. \figref{fig:struct1-ports} shows the resulting P--K--$\Delta$ interconnection.

Closing $K$ gives the lower linear fractional transformation (LFT) $N=\mathcal{F}_l(P,K)$ \cite{ZhouDoyleGlover1996}. Signals are treated as elements of $\ell_2$ and the interconnected systems as operators on $\ell_2$, so every gain below is an induced norm. Assume this nominal LTI loop is internally stable and partition it as
\begin{equation}
\label{eq:lft-partition}
  \begin{pmatrix} u_\Delta \\ z \end{pmatrix}
  = N \begin{pmatrix} y_\Delta \\ w \end{pmatrix},
  \qquad
  N=\begin{pmatrix} N_{11} & N_{12} \\ N_{21} & N_{22} \end{pmatrix}.
\end{equation}
Closing $y_{\Delta,k}=\Delta(\epsk)\,u_{\Delta,k}$ gives the upper LFT $\mathcal{F}_u(N,\Delta_k)$. The nominal return map $N_{11}:y_\Delta\mapsto u_\Delta$ determines robust stability; $N_{22}:w\mapsto z$ describes nominal performance.

\subsection{Small-Gain Certificate for Time-Varying Jitter}

Checking the closed loop for stability at each constant jitter value $|\varepsilon|\leq\epsbar$ is tempting in practice, because each such loop is LTI. However, this check does not guarantee stability when the jitter varies. The state transition over several samples is then a product of closed-loop matrices for successive jitter values. Such a product need not be Schur stable, even when every factor is. Section~\ref{sec:example} shows a controller that is stable at the constant values $\varepsilon_k\equiv\pm0.15$ but diverges when the jitter alternates between them. The certificate must therefore hold for time-varying jitter, so we use the operator form of the small-gain theorem \cite{DesoerVidyasagar1975}.

\begin{lem}[Operator small gain \cite{DesoerVidyasagar1975}]
\label{lem:small-gain}
Consider two bounded causal linear operators $H_1$ and $H_2$ on $\ell_2$, interconnected in feedback by
\[
  e_1 = v_1-H_1 e_2, \qquad
  e_2 = v_2-H_2 e_1,
\]
where $v_1,v_2$ are the exogenous inputs and $e_1,e_2$ the loop outputs of the two blocks. This interconnection is well posed and finite-gain $\ell_2$-stable from $(v_1,v_2)$ to $(e_1,e_2)$ if
\begin{equation}
\label{eq:cond-lemma}
  \|H_1\|_{\ell_2\to\ell_2}\,
  \|H_2\|_{\ell_2\to\ell_2}<1.
\end{equation}
\end{lem}

\noindent
The lemma holds for time-varying operators. In this paper, $H_1=\mathcal D$ is the jitter operator introduced in~\eqref{eq:jitter-operator} and \linebreak 
$H_2=N_{11}$ is the nominal return map of the closed loop~\eqref{eq:lft-partition}.

Because the jitter operator $\mathcal D$ is memoryless, \eqref{eq:uncertainty-bound} bounds its induced norm by $\bar\delta$. The nominal return map $N_{11}$ is stable and LTI, so its induced norm equals its $\Hinf$ norm:
\begin{equation}
\label{eq:norm-n11}
  \|N_{11}\|_{\ell_2\to\ell_2}=\Hinfnorm{N_{11}}
    =\sup_{\omega\in[-\pi,\pi]}\bar\sigma\bigl(N_{11}(e^{j\omega})\bigr).
\end{equation}
The frequency $\omega$ is the normalized discrete-time frequency, so the time-varying robustness test needs only an offline bound of~$\Delta_k$ and the frequency response of the LTI system~$N_{11}$. The operator norm condition~\eqref{eq:cond-lemma} therefore simplifies to the matrix norm condition~\eqref{eq:smallgain-condition}:

\begin{thm}[Jitter-robust stability]\label{thm:jitter-stability}
Let the nominal closed loop $N$ be internally stable. The perturbed loop $\mathcal{F}_u(N,\Delta_k)$ is well posed and finite-gain $\ell_2$-stable for every jitter sequence in $\mathcal{E}(\epsbar)$ if
\begin{equation}\label{eq:smallgain-condition}
  \bar\delta\,\Hinfnorm{N_{11}} < 1
\end{equation}
with $\bar\delta$ from \eqref{eq:uncertainty-bound} and~$\Hinfnorm{N_{11}}$ from \eqref{eq:norm-n11}.
\end{thm}

\begin{proof}
 With \eqref{eq:uncertainty-bound} and~\eqref{eq:norm-n11}, condition~\eqref{eq:smallgain-condition} implies $\|\mathcal D\|_{\ell_2\to\ell_2}\,\|N_{11}\|_{\ell_2\to\ell_2}<1$. Lemma~\ref{lem:small-gain} therefore applies uniformly over all sequences in $\mathcal{E}(\epsbar)$. The other partitions of $N$ in \eqref{eq:lft-partition} are stable, so the performance channel is also finite-gain stable.
\end{proof}

For one full uncertainty block, this test has no $D$-scale conservatism \cite{PackardDoyle1993,Shamma1994}. Conservatism enters where the one-parameter family $\{\Delta(\varepsilon)\}$ of \eqref{eq:jitter-block} is replaced by the norm-bounded set $\{\Delta\mid\|\Delta\|_2\leq\bar\delta\}$, and where all sequences in $\mathcal{E}(\epsbar)$ must be covered at once.

\subsection{From Input--Output to Uniform Exponential Stability}
\label{sec:exp-stability}

The strict condition~\eqref{eq:smallgain-condition} also yields a common quadratic Lyapunov function. This function extends the input--output certificate to \emph{uniform exponential stability} (UES) of the state.

Let $(A_N,B_N,C_N,D_N)$ be the realization of $N_{11}$ on the plant--controller state $\xi_k=[x_k\Tr,x_{K,k}\Tr]\Tr$, which retains all plant and controller modes. Internal stability of the nominal loop makes $A_N$ Schur stable.

\begin{thm}\label{thm:exp-stability}
Let the nominal closed loop be internally stable and let \eqref{eq:smallgain-condition} hold. Then the origin of the closed-loop time-varying recursion
\[
  \xi_{k+1} = \bigl(A_N + B_N\Delta(\epsk)\bigl(I-D_N\Delta(\epsk)\bigr)^{-1}C_N\bigr)\xi_k
\]
is uniformly globally exponentially stable, uniformly over all admissible jitter sequences.
\end{thm}

\begin{proof}
For $\bar\delta=0$, the loop is nominal and internally stable. Otherwise, set $\widetilde y_{\Delta,k}=y_{\Delta,k}/\bar\delta$ and $\widetilde\Delta(\epsk)=\Delta(\epsk)/\bar\delta$. The normalized return map $\widetilde N_{11}=\bar\delta N_{11}$ has realization $(A_N,\bar\delta B_N,C_N,\bar\delta D_N)$ on the same state $\xi_k$. Now $A_N$ is Schur stable, $\Hinfnorm{\widetilde N_{11}}<1$, and $\|\widetilde\Delta(\epsk)\|_2\le1$. The discrete-time bounded-real lemma \cite[Ch.~21]{ZhouDoyleGlover1996} gives \linebreak $P=P\Tr\succ0$, $\eta>0$, and $V(\xi)=\xi\Tr P\xi$ satisfying
\[
  V(\xi_{k+1})-V(\xi_k)
  \le \|\widetilde y_{\Delta,k}\|_2^2-\|u_{\Delta,k}\|_2^2-\eta\|\xi_k\|_2^2
\]
along the normalized realization, for every state and uncertainty input. One matrix $P$ therefore serves every $\Delta(\epsk)$. Since $\|\bar\delta D_N\|_2<1$, the feedback is well posed. Closing $\widetilde y_{\Delta,k}=\widetilde\Delta(\epsk) u_{\Delta,k}$ makes the supply term nonpositive, so $V(\xi_{k+1})-V(\xi_k)\le-\eta\|\xi_k\|_2^2$. The resulting common quadratic Lyapunov function establishes UES \cite[Sec.~5.1]{Boyd1994}.
\end{proof}

Theorems~\ref{thm:jitter-stability} and~\ref{thm:exp-stability} use only the bounds $\|\Delta(\epsk)\|_2\le\bar\delta$ and therefore hold for every sequence of real matrices with this bound, not only those generated by jitter.

Between samples,
\[
  x(t_k+\tau)=e^{A\tau}x(t_k)+\int_0^\tau e^{As}B u_k\,ds,
\]
with $0\le\tau\le(1+\epsbar)\ts$. The hold operators and the maps from $\xi_k$ to $x(t_k)$ and $u_k$ are uniformly bounded. Exponential decay at the sampling instants therefore extends to the continuous-time plant state. Inter-sample stability follows without an inter-sample performance specification.

\section{Workflow for Controller Synthesis}
\label{sec:tooling}

\begin{figure}[htbp]
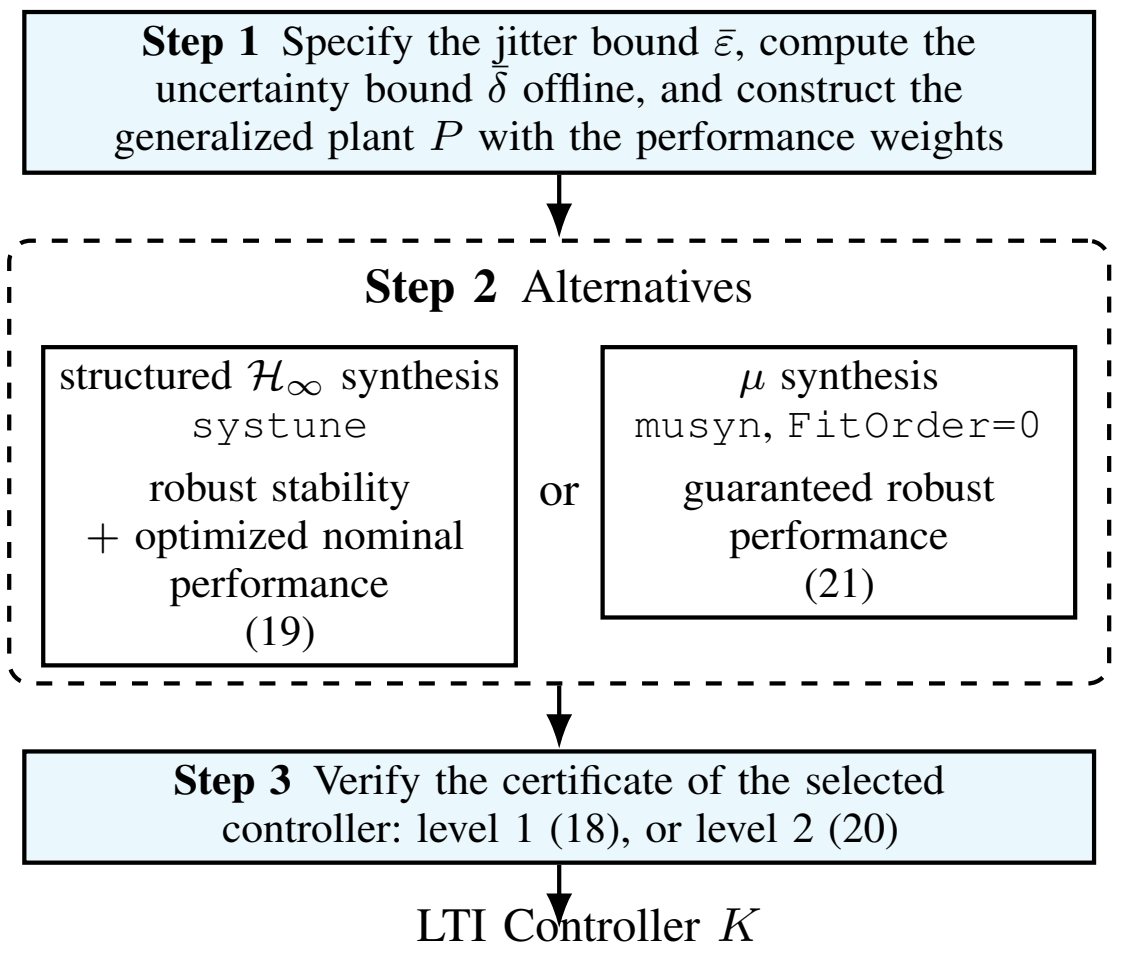

  \centering
  \begin{ctrldiagram}[node distance=0.85cm and 1.0cm]
    \node[block, fill=cyan!7, align=center, text width=6.9cm, text height=, text depth=] (step1)
      {\textbf{Step 1}\enspace Specify the jitter bound $\epsbar$, compute the uncertainty bound $\bar\delta$ offline, and construct the generalized plant $P$ with the performance weights};
    \def\arrgap{12pt}%
    \node[font=\small, anchor=north, inner sep=2pt, text height=, text depth=] (steplbl) at ($(step1.south)+(0,-15.4pt)$)
      {\textbf{Step 2}\enspace Alternatives};
    \node[block, align=center, text width=2.95cm, text height=, text depth=, anchor=north] (hinfinf) at ($(steplbl.south)+(-1.85cm,-4pt)$)
      {structured $\Hinf$ synthesis\\ \texttt{systune}\\[3pt]
       robust stability\\ $+$ optimized nominal performance\\ \eqref{eq:stability-synthesis}};
    \node[block, align=center, text width=2.95cm, text height=, text depth=, anchor=north] (musynb) at ($(steplbl.south)+(1.85cm,-4pt)$)
      {$\mu$ synthesis\\ \texttt{musyn}, \texttt{FitOrder=0}\\[3pt]
       guaranteed robust\\ performance\\ \eqref{eq:performance-synthesis}};

    \node[font=\small, text height=, text depth=] at ($(hinfinf.east)!0.5!(musynb.west)$) {or};
    \node[group-box, inner xsep=0.6em, inner ysep=3pt, fit=(hinfinf) (musynb) (steplbl)] (alt) {};

    \node[block, fill=cyan!7, align=center, text width=6.9cm, text height=, text depth=, anchor=north] (step3) at ($(alt.south)+(0,-\arrgap)$)
      {\textbf{Step 3}\enspace Verify the certificate of the selected controller: level~1~\eqref{eq:smallgain-condition}, or level~2~\eqref{eq:performance-certificate}};

    \node[output] (kout) at ($(step3.south)+(0,-\arrgap)$) {};
    \node[font=\small, anchor=mid, text height=, text depth=] at ($(kout)+(0.2cm,0)$) {LTI Controller $K$};

    \draw[signal] (step1.south) -- (alt.north);
    \draw[signal] (alt.south) -- (step3.north);
    \draw[signal] (step3.south) -- (kout);
  \end{ctrldiagram}
  \caption{Tool chain for the synthesis of jitter-robust controllers. The two alternatives of Step~2 differ in what they certify (\tabref{tab:certificates}).}
  \label{fig:workflow}
\end{figure}

The synthesis workflow has three steps:
\begin{enumerate}
\item Choose the jitter bound $\epsbar$ from a timing specification or from measurements, compute the uncertainty bound $\bar\delta$ from \eqref{eq:uncertainty-bound}, and construct the generalized plant $P$ including the performance weights.
\item Produce a fixed controller with one of the two synthesis tools: Structured $\Hinf$ synthesis yields robust stability with optimized nominal performance, and $\mu$ synthesis yields robust performance. Both take $P$ and return a fixed discrete-time $K$.
\item Verify the certificate of the selected controller with the conditions below, which differ in the property they certify (\tabref{tab:certificates}).
\end{enumerate}
\figref{fig:workflow} shows these steps and the two alternatives, and \tabref{tab:certificates} lists the certificates obtained at their end.

\subsection{Robust Stability with Structured $\Hinf$ Synthesis}
\label{sec:hinfsyn-synthesis}
For robust stability with optimized nominal performance, the design problem is
\begin{equation}\label{eq:stability-synthesis}
  \begin{aligned}
    \min_K\;&\Hinfnorm{N_{22}(K)}\\
    \text{subject to}\;&\bar\delta\,\Hinfnorm{N_{11}(K)}<1
  \end{aligned}
\end{equation}
over a prescribed controller structure with an internally stable nominal loop. Structured $\Hinf$ synthesis imposes the small-gain condition as a hard constraint \cite{ApkarianNoll2006,GahinetApkarian2011}. The constraint certifies stability against every admissible jitter sequence, while the objective optimizes nominal performance. To enforce a performance target $\gamma$ under jitter, the performance channel must enter the robustness condition itself.

\subsection{Robust Performance with $\mu$ Synthesis}
\label{sec:musyn-primer}

For a target induced gain $\gamma$, normalize the jitter operator $\mathcal D$ by $\bar\delta$ and the performance output $z$ by $\gamma$. Robust performance follows from robust stability after adding a fictitious unit-gain full block $\Delta_P$ from $z/\gamma$ to $w$ \cite[Thm.~4.3]{PackardDoyle1993}. The augmented operator is $\widehat\Delta=\mathrm{diag}(\mathcal D/\bar\delta,\Delta_P)$, and its nominal return map is
\[
  \widehat N_\gamma(K)=
  \begin{pmatrix}
    \bar\delta N_{11}(K) & N_{12}(K) \\
    \bar\delta N_{21}(K)/\gamma & N_{22}(K)/\gamma
  \end{pmatrix}.
\]
Let $n_w$ and $n_z$ be the dimensions of $w$ and $z$, respectively. Choose constant positive scalars $d_\Delta,d_P$ and port scalings $D_o,D_i$:
\[
  \begin{aligned}
    D_o&=\mathrm{diag}(d_\Delta I_{n+m},d_P I_{n_z}),\\
    D_i&=\mathrm{diag}(d_\Delta I_n,d_P I_{n_w}).
  \end{aligned}
\]
The scaled small-gain condition is
\begin{equation}\label{eq:performance-certificate}
  \Hinfnorm{D_o\widehat N_\gamma(K)D_i^{-1}}<1.
\end{equation}
Identical constant scalars on the two ports of each full block give $D_i\widehat\Delta=\widehat\Delta D_o$, even when jitter varies at every sample. Thus the scaled uncertainty remains unit-gain, and Lemma~\ref{lem:small-gain} certifies robust stability of the augmented loop. The same condition gives $\|T_{w\to z}\|_{\ell_2\to\ell_2}<\gamma$ for every admissible jitter sequence, with zero initial state \cite[Thm.~10.4]{PackardDoyle1993}.

The corresponding robust-performance synthesis objective is
\begin{equation}\label{eq:performance-synthesis}
  \begin{aligned}
    \min_{K,\gamma,d_\Delta,d_P}\;&\gamma\\
    \text{subject to}\;&\Hinfnorm{D_o\widehat N_\gamma(K)D_i^{-1}}<1
  \end{aligned}
\end{equation}
with $\gamma,d_\Delta,d_P>0$, a prescribed controller structure, and an internally stable nominal loop. Unlike \eqref{eq:stability-synthesis}, this objective bounds the weighted gain under jitter rather than only at the nominal interval.

Constant scalings keep the time-varying-jitter argument intact \cite[Thm.~10.4]{PackardDoyle1993}, because a dynamic scaling mixes samples with different jitter matrices. For a one-sample delay~$\mathcal S$ on the respective ports and $k\geq1$,
\[
  \begin{aligned}
    (\mathcal S\mathcal D u_\Delta)_k &= \Delta(\varepsilon_{k-1})\,u_{\Delta,k-1}\\
    (\mathcal D\mathcal S u_\Delta)_k &= \Delta(\epsk)\,u_{\Delta,k-1}.
  \end{aligned}
\]
These operators need not commute when jitter changes. A frequency-dependent scaling therefore cannot pass through the jitter block without changing the interconnection.

In \texttt{musyn}, the constant-scaling setting uses a zeroth-order D-fit. D--K iteration need not converge or find the global optimum of \eqref{eq:performance-synthesis}, so we check the fitted certificate for the selected controller in its chosen performance normalization.

\tabref{tab:certificates} summarizes the resulting sequence of certificates.

\begin{table}[htbp]
\centering
\caption{Sequence of certificates. Levels 1 and 2 hold for every admissible jitter sequence. Level 2 requires a fitted constant-scaling bound below one for the selected controller.}
\label{tab:certificates}
\small
\setlength{\tabcolsep}{3pt}
\begin{tabular}{>{\raggedright\arraybackslash}p{0.18\columnwidth}>{\raggedright\arraybackslash}p{0.38\columnwidth}>{\raggedright\arraybackslash}p{0.35\columnwidth}}
\hline
Level & Condition / tool & Certified property \\[0.5ex]
\hline
0: nominal & $\Hinf$ design on the jitter-free plant & nominal performance ($\epsk\equiv0$) \\[0.5ex]
\hline
1: robust stability & \eqref{eq:smallgain-condition}: hard constraint in structured $\Hinf$ synthesis, or a posteriori check & finite-gain stability \cite{DesoerVidyasagar1975}, UES (Thm.~\ref{thm:exp-stability}), nominal performance \\[0.5ex]
\hline
2: robust performance & \eqref{eq:performance-certificate}: $\mu$ synthesis, constant scaling & stability and robust performance \cite[Thms.~4.3, 10.4]{PackardDoyle1993} \\
\hline
\end{tabular}
\end{table}

Frequency-dependent $\mu$ scalings, LTI worst-case-gain and robust-stability analyses, and Bode plots do not by themselves establish the sequence-uniform certificates in \tabref{tab:certificates} \cite{Shamma1994}.

\subsection{Design of the Jitter Bound}
\label{sec:tuning}

The certificates hold for every sequence in $\mathcal{E}(\epsbar)$, even those the platform never produces. The bound $\epsbar$ can therefore be viewed as a tuning knob: Choosing a \emph{design jitter bound} $\epsdes\le\epsbar$ and
\begin{equation}\label{eq:design-bound}
  \bar\delta_{\mathrm{des}}
  = \max_{|\varepsilon|\leq\epsdes}\bar\sigma\bigl(\Delta(\varepsilon)\bigr)
  \le \bar\delta
\end{equation}
in place of $\bar\delta$ trades certified jitter robustness for performance, since a larger $\Hinfnorm{N_{11}}$ becomes admissible. This gives the engineer a degree of freedom in tuning the controller, but also moves reponsibility for the sequences outside $\mathcal{E}(\epsdes)$ to the engineer. The certificate covers $\mathcal{E}(\epsdes)$ alone, so ruling out the discarded tail $|\epsk|>\epsdes$ is left to the engineer's timing measurements.

\section{Example: Lightly Damped Second-Order Plant}\label{sec:example}

The example illustrates the distinction between nominal design, robust stability with nominal performance, and robust performance in \tabref{tab:certificates}.

\subsection{Design Setup}

The plant $G(s)=1/(s^2+0.2\,s+1)$ is lightly damped, with $n=2$ and $m=p=1$. For $\ts=0.1\,\mathrm{s}$ and $\epsbar=0.2$, the sampling interval ranges over $[0.08,0.12]\,\mathrm{s}$. The exact jitter block has size $2\times3$.
A grid of $1000$ equally spaced jitter values gives the uncertainty bound $\bar\delta=0.02832$ for the jitter bound $\epsbar=0.2$. A finite grid can miss the true maximum between grid points, so its $\bar\delta$ can sit slightly below the true one. A Lipschitz bound on the jitter block or the closed-form bound in the Appendix closes this gap. Both raise $\bar\delta$ and are omitted due to space limitation.

The tracking objective is weighted sensitivity $z=W_zv$, where $v=r-y$ is the controller input. The first-order discrete weight $W_z$ has DC gain $30$, unit-gain crossover $0.1\,\mathrm{rad/s}$, and high-frequency gain $0.1$. Following \figref{fig:struct1-ports}, the weighted plant has inputs $[y_\Delta,r,u]$ and outputs $[u_\Delta,z,v]$, with the exogenous port carrying $w=r$. For $\mu$ synthesis, its jitter ports are connected to a full norm-bounded uncertainty block with the uncertainty bound $\bar\delta$.

We use MATLAB R2026a for all computations. All controllers have order three. The \texttt{systune} run uses \texttt{rng(0)} and \texttt{RandomStart=5}; \texttt{musyn} uses \texttt{FitOrder=0}, selecting the iterate with minimum \texttt{PeakMuFit}. Other algorithmic options retain their defaults. The MATLAB scripts are available from the authors on request.

\subsection{Controller Comparison}

The three controllers share the plant, the weight $W_z$, and the controller order; they differ only in whether and how the jitter requirement enters the synthesis. The nominal $\Hinf$ controller $K_\mathrm{nom}$ optimizes the zero-jitter loop, but its small-gain product is $84.3$, far above one. The structured $\Hinf$ controller $K_\mathrm{rob}$ holds the small-gain product at $0.90$, a $10\%$ margin on \eqref{eq:smallgain-condition}, and retains nominal performance while certifying stability for every sequence in $\mathcal{E}(\epsbar)$. The constant-scaled $\mu$ controller $K_\mu$ instead includes the weighted performance channel in the uncertainty structure. Its selected fitted bound is $0.8365<1$ for $\gamma=1$, which certifies $\|T_{w\to z}\|_{\ell_2\to\ell_2}<0.84$ for every sequence in $\mathcal{E}(\epsbar)$. These checks place the designs at the three certificate levels of \tabref{tab:certificates}.

\subsection{Time-Domain Validation}

\begin{figure}[htbp]
  \centering
  \includegraphics[width=0.8\columnwidth]{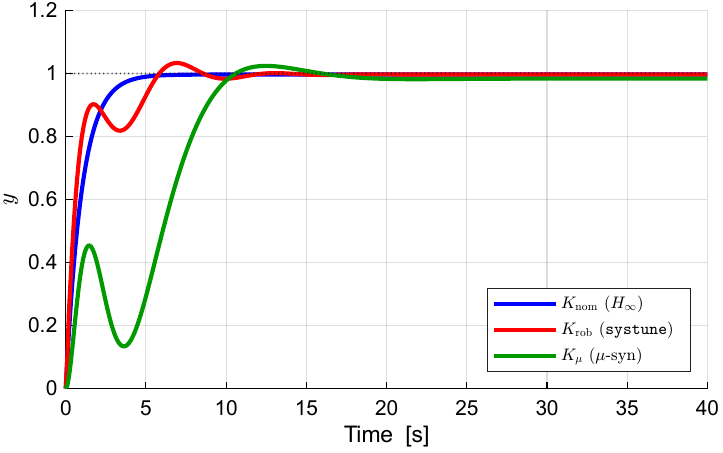}
  \caption{Nominal closed-loop step responses.}
  \label{fig:step-nominal}
\end{figure}

We simulate all cases with $|\epsk|=0.15<\epsbar=0.2$ and plot the jittered responses at the physical sampling instants $t_k$. \figref{fig:step-nominal} shows similar nominal responses for $K_\mathrm{nom}$ and $K_\mathrm{rob}$. The $K_\mu$ response is slower because the robust-performance specification forces a lower bandwidth.

\begin{figure}[htbp]
  \centering
  \begin{subfigure}{\columnwidth}
    \centering
    \includegraphics[width=0.7\columnwidth]{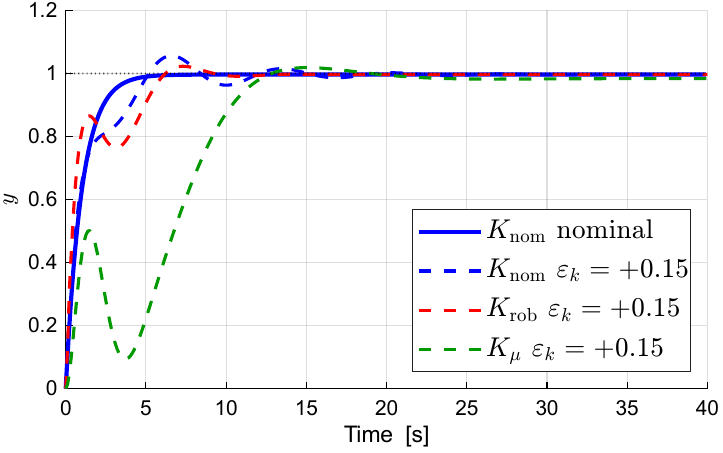}
    \caption{$\varepsilon_k\equiv+0.15$.}
    \label{fig:sim-const-pos}
  \end{subfigure}
  \par\medskip
  \begin{subfigure}{\columnwidth}
    \centering
    \includegraphics[width=0.7\columnwidth]{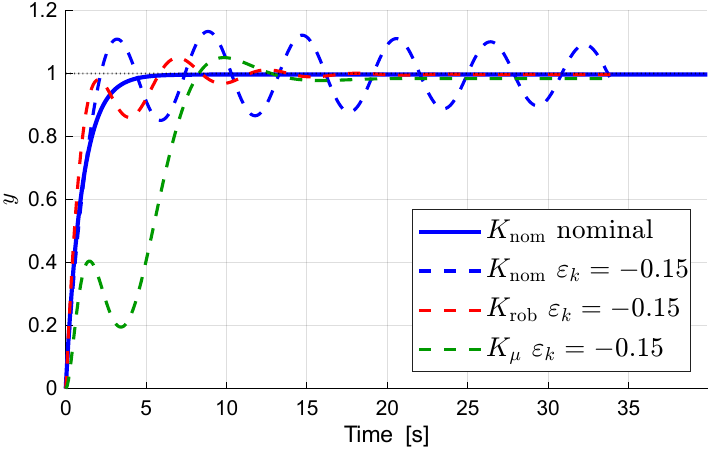}
    \caption{$\varepsilon_k\equiv-0.15$.}
    \label{fig:sim-const-neg}
  \end{subfigure}
  \caption{Step responses at constant jitter.}
  \label{fig:sim-const}
\end{figure}

\begin{figure}[htbp]
  \centering
  \begin{subfigure}{\columnwidth}
    \centering
    \includegraphics[width=0.7\columnwidth]{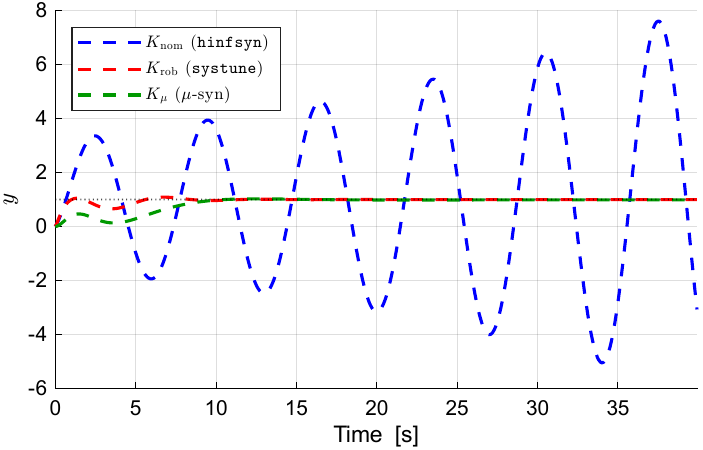}
    \caption{$K_\mathrm{nom}$ diverges; both robust controllers remain bounded.}
    \label{fig:sim-alt1a}
  \end{subfigure}
  \par\medskip
  \begin{subfigure}{\columnwidth}
    \centering
    \includegraphics[width=0.7\columnwidth]{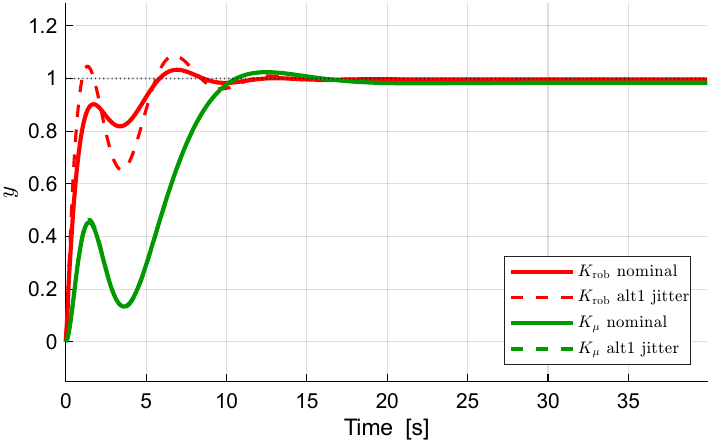}
    \caption{Zoom of the robust-controller responses.}
    \label{fig:sim-alt1b}
  \end{subfigure}
  \caption{Alternating jitter, $\varepsilon_k=\pm0.15$. Solid: zero jitter; dashed: jittered.}
  \label{fig:sim-alt1}
\end{figure}

All three controllers remain stable if the jitter remains constant at one of the two extreme values $\varepsilon_k\equiv\pm0.15$ (\figref{fig:sim-const}). When the jitter instead alternates between the two extreme values $\pm0.15$ (i.e., $\varepsilon_k=0.15 \cdot (-1)^k$ for all $k$), the loop with $K_\mathrm{nom}$ diverges whereas the loops with both robust controllers remain stable (\figref{fig:sim-alt1}). The $K_\mu$ trajectory stays closer to its zero-jitter response than the $K_\mathrm{rob}$ trajectory.

To see why the analysis with constant jitter values do not suffice, let $\xi_k=[x_k\Tr,x_{K,k}\Tr]\Tr$ and write the zero-input recursion as $\xi_{k+1}=A_{\mathrm{cl}}(\epsk)\xi_k$, where $A_{\mathrm{cl}}(\varepsilon)$ is the closed-loop state matrix with the jitter frozen at $\varepsilon$. A cycle starting with $+\varepsilon$ has the two-sample map
\[
  \xi_{k+2}=A_{\mathrm{cl}}(-\varepsilon)
             A_{\mathrm{cl}}(+\varepsilon)\xi_k.
\]
Testing constant jitter values checks each factor separately. The periodic loop instead depends on repeated powers of $A_{\mathrm{cl}}(-\varepsilon)A_{\mathrm{cl}}(+\varepsilon)$, which need not be Schur stable. By Theorem~\ref{thm:exp-stability}, the small-gain certificate yields one quadratic Lyapunov function that decreases for every admissible interval, hence for every product of the matrices $A_{\mathrm{cl}}(\varepsilon_k)$: Switching cannot destabilize the loop, and no jitter pattern need be specified.

The simulations show the distinction between the certificates: A constant-jitter check does not prevent instability under alternating jitter, and robust stability by itself leaves performance degradation untouched. The plotted sequence is not a claimed worst case. The $\mu$ design trades nominal speed for a guarantee on weighted performance over the whole jitter envelope $\mathcal{E}(\epsbar)$. Lower bandwidth could reduce jitter sensitivity on its own. The certificate is what distinguishes $K_\mu$ from a merely detuned design. For the robust designs, the certificate is moderately conservative: The small-gain test certifies $K_\mathrm{rob}$ up to $\epsbar=0.22$ and $K_\mu$ up to $0.26$, whereas a search over periodic jitter sequences finds instability only at $|\epsk|=0.415$ and $0.58$, respectively.

\section{Conclusion and Outlook}
\label{sec:conclusion}

The paper has shown that standard robust-control tools apply to control loops with jitter if a known jitter bound~$\epsbar$ limits the sample-to-sample jitter. This result follows from a decomposition in which the uncertainty block is memoryless. Structured $\Hinf$ synthesis then optimizes nominal performance under a jitter-stability constraint, and constant-scaled $\mu$ synthesis certifies robust performance as well. The guarantees hold for every sequence in the jitter envelope $\mathcal{E}(\epsbar)$, including uniform exponential stability.

The full block, together with the requirement to cover every sequence, is conservative relative to physical jitter. Measured stochastic jitter models and explicit inter-sample performance analysis would relax the design.
The certificate covers jitter only. Combining jitter with plant uncertainty requires a joint uncertainty model and a new certificate.

\appendix

\section{Analytical Uncertainty Bound}
\label{sec:analytical-bounds}

Submultiplicativity, $\|e^{\varepsilon A\ts}-I\|_2\le e^{|\varepsilon|\|A\|_2\ts}-1$, and the integral triangle inequality give
\begin{equation}
\begin{aligned}
\|\DeltaA{\epsk}\|_2
  &\leq \bigl(e^{\epsbar\|A\|_2\ts}-1\bigr)\|\Adnom\|_2
  =:\bar\delta_A,\\
\|\DeltaB{\epsk}\|_2
  &\leq \|B\|_2\int_{\ts}^{(1+\epsbar)\ts}e^{\|A\|_2\tau}\,d\tau
  =:\bar\delta_B.
\end{aligned}
\end{equation}
The positive-jitter integration interval bounds either sign, since the scalar majorant $\tau\mapsto e^{\|A\|_2\tau}$ is increasing and each tail interval has length at most $\epsbar\ts$. For $\|A\|_2>0$,
\begin{equation}
  \bar\delta_B=
  \frac{\|B\|_2}{\|A\|_2}\,e^{\|A\|_2\ts}
  \bigl(e^{\epsbar\|A\|_2\ts}-1\bigr).
\end{equation}
At $A=0$, $\bar\delta_B=\epsbar\ts\|B\|_2$. Cauchy--Schwarz gives the full-block bound $\|\Delta(\epsk)\|_2\leq\sqrt{\bar\delta_A^2+\bar\delta_B^2}=:\bar\delta$.

\bibliographystyle{IEEEtran}
\bibliography{Literature}

\end{document}